\documentclass[12pt]{iopart}

\usepackage{iopams}
\usepackage{cite}

\usepackage{amssymb}
\usepackage{braket} 
\usepackage{svg}
\usepackage{graphicx}
\usepackage{xcolor}

\definecolor{Luk1}{rgb}{0.7,0.35,0}

\begin{document}

\title[Quantum Synchronization in Presence of Shot Noise]{Quantum Synchronization in Presence of Shot Noise}

\author{
Florian H\"ohe$^1$,
Lukas Danner$^{1,2}$, 
Ciprian Padurariu$^1$,
Brecht I. C Donvil$^1$, 
Joachim Ankerhold$^1$, and 
Bj\"orn Kubala$^{1,2}$
}

\address{$^1$Institute for Complex Quantum Systems and IQST, University of Ulm, 89069 Ulm, Germany}
\address{$^2$Institute for Quantum Technologies, German Aerospace Center (DLR), 89081 Ulm, Germany}
\ead{florian.hoehe@uni-ulm.de}
\vspace{10pt}

\begin{abstract}
Synchronization is a widespread phenomenon encountered in many natural and engineered systems with nonlinear classical dynamics. How synchronization concepts and mechanisms transfer to the quantum realm and whether features are universal or platform specific are timely questions of fundamental interest. 
They can be studied in superconducting electrical circuits which provide a well-established platform for nonlinear quantum dynamics. 
Here, we consider a Josephson-photonics device, where a dc-biased Josephson junction creates (non-classical) light in a microwave cavity. The combined quantum compound constitutes a self-sustained oscillator: a system susceptible to synchronization. This is due to the inherent effect of an in-series resistance, which realizes an autonomous feedback mechanism of the charge transport on the driving voltage. Accounting for the full counting statistics of transported charge not only yields phase diffusion, but allows us to describe phase locking to an ac-signal and the mutual synchronization of two such devices.
Thereby one can observe phase stabilization leading to a sharp emission spectrum as well as unique charge transport statistics revealing shot noise induced phase slips. Two-time perturbation theory is used to obtain a reduced description of the oscillators phase dynamics in form of a Fokker-Planck equation in generalization of classical synchronization theories. 
\end{abstract}

%
%

\section{Introduction}
Since Huygens's first study of  pendulum clocks in 1665 \cite{Huygens} the nonlinear phenomenon of synchronization has  extensively been studied in all fields of science \cite{Pikovsky2001,strogatz2000,kuramoto2003,strogatz1993,glass2001}. It occurs in systems autonomously undergoing \emph{self-sustained oscillations}, i.e., these are not powered by an external periodic drive, but are kept going by resupplying energy in an incoherent manner, such as by the gravity-driven weight in a pendulum clock or by a battery. In the absence of any periodic reference oscillation, self-sustained oscillations  do not have any preferred phase which instead will be randomized by noise acting on the system.
In the same manner, the phase will be strongly susceptible to regular external perturbations, so that even a weak injected signal can entrain one oscillator or a weak coupling can synchronize the motion of two such systems. While the oscillation spectrum of the free self-sustained oscillator is broad and centered around its natural frequency, it will be pulled towards the frequency of the injected signal and become very sharp when locked, explaining the technological relevance of locking and synchronization, e.\,g. in laser physics \cite{BuczekIEEE173,vanTartwijk_PQE_1998}.

Unsurprisingly, also quantum synchronization has attracted interest recently, both for systems whose classical counterparts show synchronization behavior with the paradigmatic example of the van der Pol oscillator \cite{Walter2014,Amitai2017,walter2015,Lee2013,weiss2017,lee2014,Dutta2019,Arosh2021}, as well as for systems \cite{Roulet2018lock,Para2020,Koppenhofer2020,Roulet2018synch,jaseem2020, Hush2015, Laskar2020, krithika2022,Zhang_PRR_2023}, such as small spins, or mechanisms \cite{es2020,sonar2018,Loerch2017,nigg2018,Shen_PRE_2023,Lutz_PRL_2024,kehrer2024} 
without classical analogue.

Superconducting electrical circuits incorporating Josephson junctions are an ideal platform to study, both, the fundamental physics of (circuit) quantum electrodynamics as well as the technological exploitation of nonlinear quantum dynamics. For instance, such circuits can oftentimes bridge 
all the way from the semiclassical to the deep quantum regime. One manifestation of quantum effects is the appearance of shot noise: fluctuations of the electrical current associated with the granular nature of electrical charges, i.e., of electrons in a metal or (Cooper-)pairs of electrons in a superconductor. Synchronization physics in such devices is thus influenced by the quantum mechanical nature of the nonlinear oscillator and the quantum stochastic nature of the noise affecting its phase.

A perfect system to study such quantum synchronization in the presence of shot noise are Josephson-photonics devices, where Cooper pairs (CPs) tunneling across a dc-biased Josephson junction create photons in an in-series microwave cavity \cite{Hofheinz2011}.
In such devices, the nonlinearity of the Josephson junction imprints a quantum nature on the microwave generation so that they can be employed to create entangled light \cite{Peugeot2021}, photon multiplets \cite{Menard2022}, or single-photons \cite{Grimm2019,Rolland2019}. Constituting versatile and bright quantum light sources they may become an important component for quantum technological applications \cite{casariego2022,gu2017} in quantum communication, quantum information processing, and for sensing and imaging tasks, such as envisioned quantum radars \cite{chang2019}. While synchronization in (multi-mode) Josephson photonics has been observed in a semiclassical regime \cite{Cassidy2017,Chen2014,Yan2021} and to some extent theoretically described \cite{Simon2018,Danner2021,Lang2023}, those devices are also particularly suited to study the deep quantum regimes of strong nonlinearity at the few-photon level \cite{Rolland2019,Menard2022,Gramich2013}, where such fundamental questions as quantum measures of synchronization \cite{Mari2013,ameri2015,shen_Entropy_2023} and synchronization of individual trajectories \cite{Lutz_PRL_2024,es2020} can be studied.

Here, we present a model for the quantum dynamics of a Josephson-photonics circuit biased by a realistic voltage source. We show that the system is driven into self-sustained oscillations and describe how it synchronizes to an external ac-signal or a second circuit. Such phase locking enables an immediate exploitation of these devices as sources of entanglement, thus dispensing with the elaborate scheme developed in \cite{Peugeot2021}  to characterize the entanglement of a two-mode squeezed state without phase stability, and opens the door to Wigner state tomography consequently broadening the potential technological impact of these devices. Moreover, this new platform for quantum synchronization allows studying fundamental limitations of phase stabilization in a regime, where the quantum shot noise of charge tunneling-events feeds back into the Hamiltonian and induces phase slips.

\begin{figure}[b]
\includegraphics[width=\textwidth]{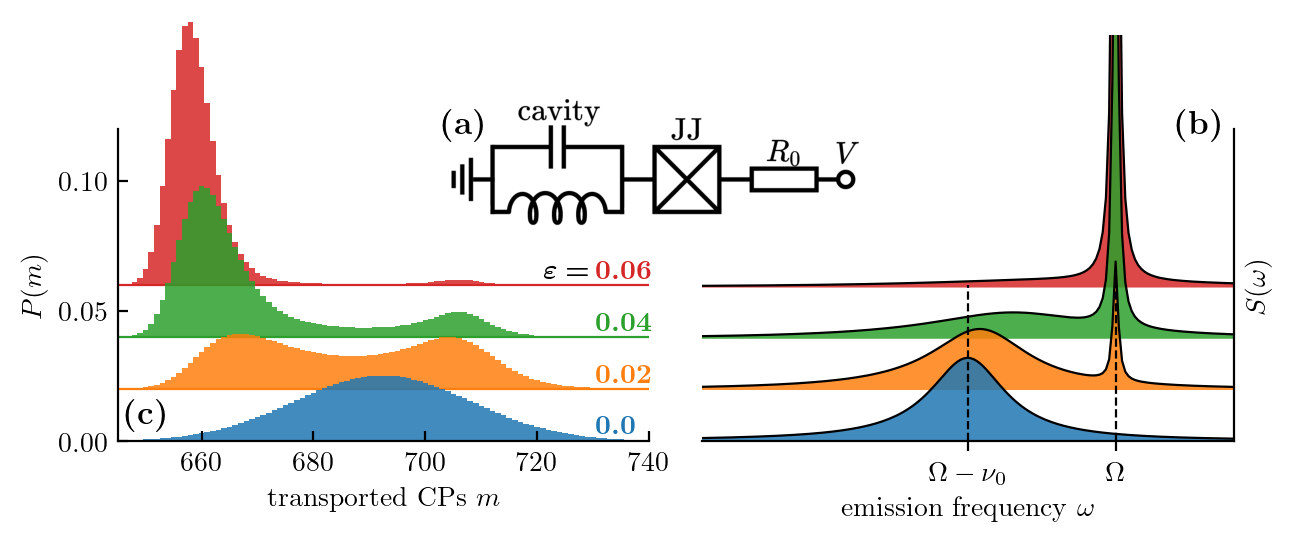}
\caption{The microwave emission of a Josephson-photonics device (a), consisting of a Josephson junction in-series to a microwave cavity and a resistance $R_0$, can be locked by adding a small ac-signal (of  strength $\varepsilon$ and frequency $\Omega$ on top of a dc-voltage drive). (b) Emission spectra $S(\omega)$ are centered around a frequency $\omega_J$ determined by the voltage across the junction but broadened by phase fluctuations. Applying a locking signal [initially detuned by $\nu_0=\Omega - \omega_J(\varepsilon=0)\,$] pulls the emission frequency towards $\Omega$, eventually resulting in a sharp emission peak at $\Omega$ [here artificially broadened by $\gamma/1000$ (FWHM)]. (c) The counting statistics of CPs transported within a time interval $T=267/\gamma$ is directly linked to the statistics of voltage fluctuations and thus the spectral width.
Increasing the locking strength narrows the CP distribution, while shot noise induces phase slips by $\Delta m=1/r_0 $. 
[Parameters: $E_J^*/(\hbar\gamma)=20$, $\alpha_0=0.1$, $2eV_\mathrm{dc}/\hbar=\omega_0$, $\nu_0 = 1.25 r_0\gamma$, $r_0 = 1/50$.]}
\label{fig:SpectrumWaterfall}
\end{figure}

\section{Quantum synchronization to an external signal}
In a Josephson-photonics device Cooper pairs tunneling across a dc-biased Josephson junction create photonic excitations in a microwave cavity connected in series with the junction, see figure~\ref{fig:SpectrumWaterfall}(a). The energy of Cooper pairs is converted 
into microwave radiation emitted from the leaky cavity with a frequency, $\omega_J \approx 2e V_{\mathrm{dc}}/\hbar = \omega_\mathrm{dc}$, when the dc-voltage is tuned close to a resonance frequency, $\omega_0 \approx \omega_J$, of the cavity.
Placing a resistor $R_0$ in series with junction and cavity promotes a back-action of the Cooper-pair tunneling 
on the voltage driving the tunneling process: the mean dc-current $\langle \hat{I}_\mathrm{CP} \rangle$ reduces the effective dc-voltage driving the junction, $\omega_J = 2e (V_{\mathrm{dc}} -R_0 \langle \hat{I}_\mathrm{CP} \rangle)/\hbar$, while current shot noise is turned into voltage fluctuations assumed here to exceed thermal or other fluctuations of the voltage source.
As we demonstrate in the following, the very same backaction mechanism can help to avert broadening, if a small ac-locking signal is added to the dc-voltage,  $V(t) = V_{\mathrm{dc}} + V_{\mathrm{ac}} \cos[\Omega t + \varphi_{\mathrm{ac}}]$. Increasing the amplitude $\varepsilon = 2e V_{\mathrm{ac}}/(\hbar \Omega)$, the spectrum will first show an additional small $\delta$-peak at $\Omega$, while the broadened peak is pulled towards $\Omega$, until eventually most light is emitted within a very sharp peak in sync with the external locking drive, see figure~\ref{fig:SpectrumWaterfall}(b). How to model and simulate these fundamental features of nonlinear quantum dynamics \--- namely frequency pulling, phase locking, and the 
closely related synchronization to a second device, which takes the role of the external locking signal 
\--- is the central result of this work.

\subsection{Model}
A Josephson-photonics device is modeled by the Hamiltonian
\begin{equation}
 \label{eq:HamiltonianJJPhot_Time}
 H_S = \hbar \omega_0 \hat{a}^{\dagger}\hat{a} - E_J \cos\left[\alpha_0 (\hat{a}^{\dagger} + \hat{a}) + \frac{2e}{\hbar}\int_0^t V(\tau) \mathrm{d}\tau - \hat{\varphi}_{R_0} \right] \, ,
\end{equation}
where $\hat{a}^\dagger$ and $\hat{a}$ denote creation and annihilation operators of the cavity with zero-point fluctuations $\alpha_0 = (2e^2 \sqrt{L/C}  /\hbar)^{1/2}$, $E_J$ is the Josephson energy, and the argument of the cosine follows from Kirchhoff's sum rule. In contrast to ac-driven Josephson microwave devices, such as Josephson parametric amplifiers \cite{gu2017,casariego2022}, the nonlinear Josephson-photonics device is not driven by a periodic force, but by an incoherent power input from a battery. 
It thus emulates 
the textbook 
dynamical system susceptible to locking and synchronization, the pendulum clock driven by a weight. This aspect of Josephson-photonics devices is hidden in the standard description, when a perfect dc-voltage bias $V \equiv V_\mathrm{dc}$ 
without any resistance is assumed, i.e. $\hat{\varphi}_{R_0}=0$, so that a periodic drive term $E_J \cos \left[ 2e V_\mathrm{dc} t/\hbar  + \alpha_0 (\hat{a}^{\dagger} + \hat{a}) \right]$ results. 
The oftentimes neglected fluctuations of the voltage \cite{Wang2017} turn the system into a (lockable) self-sustained oscillator. These fluctuations arise from thermal or external sources, but crucially contain an unavoidable contribution from the shot noise of the current through the junction. 
We describe this by considering a resistance in-series with a perfect voltage source, as visualized in figure~\ref{fig:SpectrumWaterfall}(a). The integrated voltage fluctuations due to shot noise on this resistance yield the phase, 
\begin{equation}
\label{eq:phase_def}
\hat{\varphi}_{R_0} = (2e/\hbar) \int_0^t \hat{V}_{R_0}\mathrm{d}\tau  = (2e/\hbar) R_0 \int_0^t \hat{I}_\mathrm{CP} \mathrm{d}\tau\,,
\end{equation}
appearing in (\ref{eq:HamiltonianJJPhot_Time}).

In general, the current in (\ref{eq:phase_def}) is a full quantum operator and so is the integral for the phase, which being proportional to the the number of Cooper-pairs $m$ transferred across the junction in the elapsed time interval incorporates the effects of coherences between different charge transfer numbers. In a model without in-series resistance these coherences are limited by the typical number of photons in the cavity and a resistance will introduce  additional decoherence. Instead of retaining the full quantum character of $\hat{\varphi}_{R_0}$ and a full microscopic model of the resistance in the Hamiltonian, we can effectively model the resistor phase by treating the number $m$ of transferred Cooper pairs as a stochastic variable with statistics described by the full counting statistics of the coherent Cooper pair tunneling process. 
Such a modeling of the driving phase in the Hamiltonian  (\ref{eq:HamiltonianJJPhot_Time}) by
an $m$-dependent c-number, $\varphi_{R_0} = 2\pi r_0 m$, where $r_0=R_0/R_Q$ is the in-series resistance in units of the superconducting resistance quantum $R_Q = h/(4e^2)$, is valid for small $r_0$, where any residual coherences between charge states would correspond to minute changes of the driving phase. 
Using the $m$-dependent Hamiltonian (\ref{eq:HamiltonianJJPhot_Time}) within a quantum master equation for the state of the system, which is conditioned on the number of tunneled Cooper pairs, resembles a feedback-based locking scheme, where the in-series resistance realizes an inherent autonomous adaptation of the driving phase to a measurement outcome. 

\subsection{Locking Hamiltonian}
\emph{Technically}, we first move to a frame rotating with the frequency $\Omega$ of the ac-signal (introducing a detuning $\Delta = \delta_\mathrm{dc} + \delta_\mathrm{ac} = (\omega_0 - \omega_\mathrm{dc}) + (\omega_\mathrm{dc} - \Omega)$) while neglecting fast oscillating terms in a rotating wave approximation (RWA). The resulting Hamiltonian (cf. \ref{sec::Ap_RWA_Ham}), $H_\mathrm{RWA} = \hbar \Delta a^\dag a + h + h^\dag$, with 
\begin{eqnarray}
    \label{eq:H_RWA}
    \fl
    h^\dag(m) &= \frac{E_J^* \alpha_0}{2} e^{i \psi} : i a^\dag  \frac{J_1\left(2\sqrt{\alpha_0^2 a^\dag a}\right)}{\sqrt{ \alpha_0^2 a^\dag a}}: 
    + \varepsilon \frac{E_J^*}{4} e^{i \psi} : a^{\dag 2} \frac{J_2\left(2\sqrt{\alpha_0^2a^\dag a}\right)}{a^\dag a} + J_0\left(2\sqrt{\alpha_0^2 a^\dag a}\right) : \nonumber \\
    \fl & \approx 
    \frac{E_J^* \alpha_0}{2} e^{i \psi} \; i a^\dag 
    +\varepsilon \frac{E_J^*}{4} e^{i \psi} \left[ \frac{\alpha_0^2 a^{\dag 2}}{2} +  \left( 1- \alpha_0^2 a^\dag a \right) \right]   \;,
\end{eqnarray}
depends on the phase-difference 
\begin{equation}
    \label{eq:psi}
    \psi(t, m) = \Omega t + \varphi_\mathrm{ac} - (\omega_\mathrm{dc} t - 2\pi r_0 m)
\end{equation}
between the ac-signal and the cavity oscillation. Here, the Josephson energy is renormalized, $E_J^* = E_J e^{-\alpha_0^2/2}$, and colons signal normal-ordering of the operators. The second line, valid for small zero-point fluctuations and weak driving, so that all Bessel functions may be expanded, allows easy interpretation of the various terms: the first term describes direct driving of the cavity by the dc-bias, where each tunneling Cooper pair creates a photon; the second term, $\propto a^{\dagger 2} \cdot E_J \cdot \varepsilon$, accounts for the generation of two photons by one CP and one ac-drive excitation; and the last term represents one CP tunneling and emitting an excitation into the drive renormalized by the presence of cavity photons (see also \ref{sec::Ap_RWA_Ham}).
The various Cooper pair transfer processes come with different leading powers of the zero-point fluctuations, $\alpha_0$, and accordingly for small $\alpha_0$ the first term of (\ref{eq:H_RWA}) is dominant (and the approximation leading to the second line is valid). In consequence, if  $\psi$ were to be a fixed external parameter, the photonic state of the cavity would have a trivial dependence on $\psi$, which essentially rotates the state space, with only minor corrections stemming from the second term. Without a locking signal, $\varepsilon = 0$, the photonic state also determines the current, which is independent of the value of $\psi$, as seen by invoking energy conservation and the photon loss rate, $\propto \gamma \langle a^\dagger a \rangle$ \cite{Gramich2013}, or by evaluating the explicitly $\psi$-dependent current operator, $\hat{I}_\mathrm{CP} = -i 2e (h^\dag - h)/\hbar$, acting on the trivially $\psi$-dependent cavity state.
The fact, that the current describes changes of the number of transferred Cooper pairs $m$ and thus of $\psi$, then leads to a driving phase that drifts and diffuses and to the self-sustained oscillator dynamics described above, cf. figure \ref{fig:PhaseSpaceDistributions}\,(a).

Turning on an oscillating locking signal, $\varepsilon > 0$, breaks the phase symmetry of such an oscillator and is generically expected to eventually lead to a state locked to specific constant values of $\psi \;\textrm{mod} \; 2\pi$. Here, the locking mechanism can be ascribed to the third CP transfer process discussed above, $\propto \varepsilon E_J \cdot (1- \alpha_0^2 a^\dagger a)$, which \-- while irrelevant in the Hamiltonian (\ref{eq:H_RWA}) as an energetic shift in the $\alpha_0 \rightarrow 0$ limit \-- yields a Shapiro-like term in the current operator and results in a $\psi$ dependence of the expectation value of the current. In particular, the $\psi$-dependence of the CP transfer and thus of $m$ in (\ref{eq:psi}) can lead to locked solutions, where $\textrm{d}\psi/\textrm{d}t =0$ for specific values of $\psi \;\textrm{mod} \; 2\pi$. 

\subsection{Time-evolution}
The locking mechanism relies on the feedback of the number of transferred Cooper pairs into the phase $\psi$ appearing in the Hamiltonian (\ref{eq:H_RWA}). We can capture this physics 
by a slight modification of the number-resolved master equation technique \cite{Xu2013}, which is routinely used for the counting of \emph{incoherent} processes, such as photon emission or other rate processes appearing as Lindblad terms in the time-evolution. To generalize to the counting of \emph{coherent} Cooper-pair transport processes, the density matrix, $\rho(t) = \displaystyle{\sum_{m}} \rho_m(t)$ ($m$ are half-integers)
, is divided into components according to the number of tunneled CPs $m$ having crossed the Josephson junction after a time $t$. 
As explained in \ref{sec::Ap_coherent_counting} the half-integer indexing accounts for the coherent nature of the transfer process. 
The components are then coupled by CP tunneling amplitudes 
\begin{eqnarray}
    \label{eq:qme}
    \fl \; \; 
    \dot{\rho}_m = \mathcal{L}_0 \rho_m-\frac{i}{\hbar}\big[&h^\dag(m-1/2) \rho_{m-1/2}+ h(m+1/2) \rho_{m+1/2} \nonumber \\
    &- \rho_{m+1/2}h^\dag(m+1/2) - \rho_{m-1/2}h(m-1/2)\big].
\end{eqnarray}
The contribution from $\mathcal{L}_0$ which includes photon loss  $\gamma\mathcal{D}[a]\rho = \gamma (a \rho a^\dag - \{a^\dag a, \rho\}/2)$ with rate $\gamma$
\begin{equation}
    \mathcal{L}_0 \rho_m = - i [\Delta a^\dag a, \rho_m] + \gamma \mathcal{D}[a]\rho_m \,,
\end{equation}
does not correspond to CP tunneling-events.
Different components $\rho_{m\pm 1/2}$ appear in a manner completely analogous to how counting fields are added with different signs to coherent forward and backward tunneling terms \cite{Novotny2013,Kubala2020}. 

Since the $m$-dependence of $h(m)$ is periodic, there exists a natural compactification, when we chose (without loss of generality) $M=1/r_0 \in \mathbb{N}$ and set $\rho_{M} = \rho_0$. This enables 
numerical simulations up to such a long time, that a very large number of CPs has been transported.

With this method (and the quantum regression theorem) the steady-state spectra of figure~\ref{fig:SpectrumWaterfall}(b) discussed above were found.
These can be compared to the counting statistics of CPs, $P(m) = \mathrm{tr}\rho_m$ after a finite time $T$, shown in figure~\ref{fig:SpectrumWaterfall}(c). When the ac-signal is absent, the distribution of counted CPs is broad, conforming with a broad emission spectrum of the cavity. An increase of the strength of the external locking signal reduces the width in the statistics of counted CPs, pulls the frequency emission towards the locking frequency, and eventually results in a sharp photon emission spectrum and a sharpened probability distribution $P(m)$. In the long-time limit the peak of the locked distribution [red in \ref{fig:SpectrumWaterfall}(c)] will not become wider, but the distribution will develop many sidepeaks, see \ref{sec::Ap_coherent_counting}, which as we will show in the next section, are associated with slips of the phase variable by $2\pi$.

\section{Phase diffusion and its constraint by locking}
\begin{figure}[b]
\includegraphics[width=\textwidth]{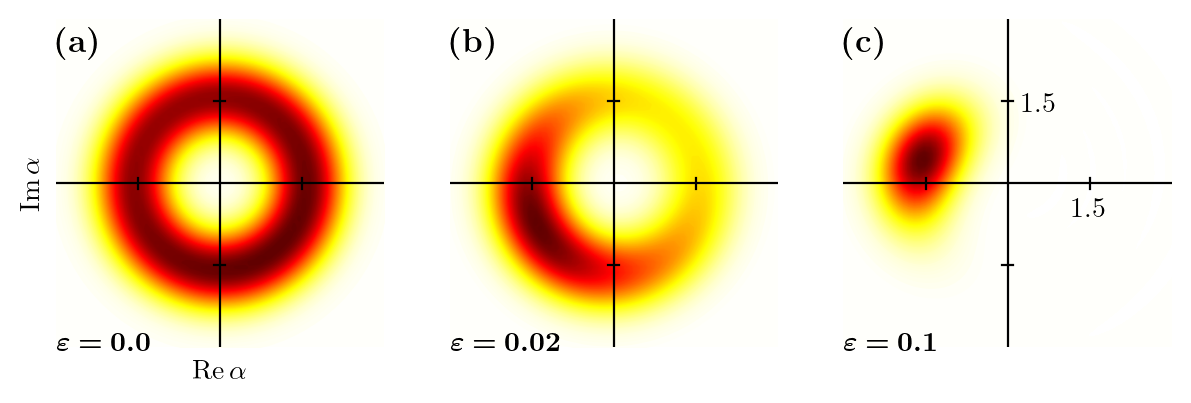}
\caption{Wigner phase-space distributions of the cavity in the steady-state. (a) Without locking signal the shot noise diffuses the oscillator's phase which becomes completely undefined. (c) The injection of a small ac-signal reduces the phase uncertainty, yielding a state with a sharply refocused phase in the locked regime. (b) In the intermediate regime shot noise and locking signal compete resulting in a state with partially diffused phase. [Parameters: $E_J^*/(\hbar\gamma)=20$, $\alpha_0=0.1$, $2e V_\mathrm{dc}/\hbar=\omega_0$, $\nu_0 = 1.25r_0\gamma$, $r_0 = 1/50$. Wigner functions are normalized to their maxima.]}
\label{fig:PhaseSpaceDistributions}
\end{figure}

\begin{figure}[b]
\includegraphics[width=\textwidth]{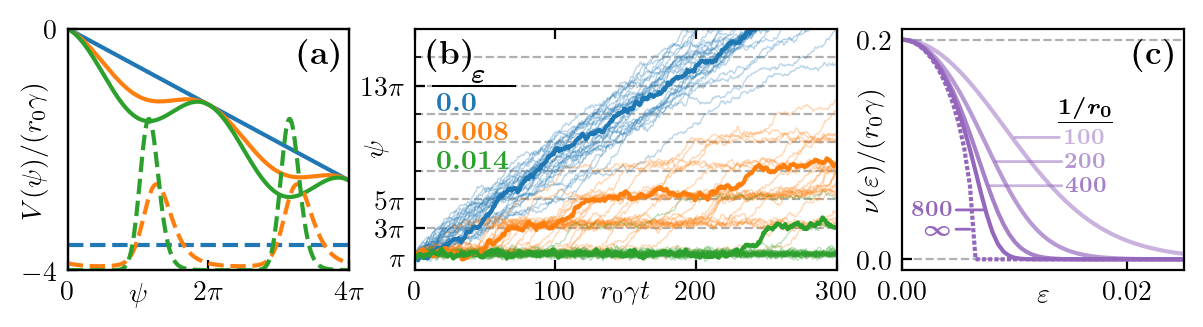}
\caption{Dynamics of the reduced phase in a tilted washboard potential. 
(a) The injection of a sufficiently strong ac-signal creates a potential with minima in which periodic steady-state distributions (dashed) are found.
(b) Individual trajectories of a phase particle in the potentials of (a). Long intervals of constant phase in the synchronized state (green) are interrupted by phase slips of $2\pi$, which proliferate (orange) and overlap (blue) in the unsynchronized regime. (c) The steady-states of (a) yield the rate $\nu(\varepsilon) = \braket{\dot{\psi}}$ of these slips. For smaller resistance the locking transition becomes more pronounced until for $1/r_0\to \infty$ the criticality of classical synchronization is recovered.
 [Parameters: $E_J^*/(\hbar\gamma)=10$, $\alpha_0=0.1$, 
 $2e V_\mathrm{dc}/\hbar=\omega_0$, $\nu_0 = 0.2r_0\gamma$, $r_0 = 1/300$.]}
\label{fig:PhaseSlips}
\end{figure}

The impact of the in-series resistance and resulting voltage fluctuations, and the effects of the locking signal can also be studied in a phase-space picture of the cavity dynamics. In figure~\ref{fig:PhaseSpaceDistributions}, quantum master equation results for the steady-state Wigner distribution are shown for different amplitudes of an injected ac-signal. The dc-voltage fixed close to resonance drives the cavity quickly into a (near coherent) state with finite amplitude. 
However, due to the lack of a reference phase, the imperfect dc-voltage drive across the junction disperses the phase of this state, until in the steady state of the system all information about the driving phase is lost [figure~\ref{fig:PhaseSpaceDistributions}(a)]. The unstable phase-space angle is mirrored by the noise-induced broadening of the emission spectrum, cf. blue spectrum in figure~\ref{fig:SpectrumWaterfall}(b). 
Turning on and progressively increasing the amplitude of the ac-voltage, the completely dephased steady-state distribution contracts [figure~\ref{fig:PhaseSpaceDistributions}(b-c)]. Concomitantly the variance of the phase-space angle is reduced and a sharpened peak emerges in the emission spectrum [cf. figure~\ref{fig:SpectrumWaterfall}(b)].
Notably, in phase-space the effects of synchronization emerge completely smoothly with increasing strength $\varepsilon$, while the spectral pulling shows remnants of the classical criticality, see figure~\ref{fig:SpectrumWaterfall}(b)] and figure~\ref{fig:PhaseSlips}(c) below. 

We will understand and quantify the observed dynamics of the phase-space angle by linking it to the notion of constrained diffusion. For that purpose, we again turn to the phase difference $\psi$ between the phase of the locking signal 
and the phase 
which  drives the cavity (\ref{eq:psi}). 
The synchronized state will then be characterized by the phase difference $\psi$  
becoming constant in time (modulo $2\pi$), as the cavity emits photons with frequencies centered sharply about the injection frequency $\Omega$.

\section{Reduced dynamics of the phase}
The question that now arises is, whether a reduced equation for the dynamics of the phase variable $\psi$ can be derived in the quantum domain, thus generalizing the Adler-theory of classical synchronization dynamics \cite{Adler1964,Pikovsky2001,Danner2021}. 
Indeed, the quantum master equation~(\ref{eq:qme}) can be rewritten as an equation for a density $\rho'(t,\psi)$ by first treating $m$ as continuous and transforming to a moving frame as shown in \ref{sec: Ap_TTPT}. Crucially, the explicit time dependence of the Hamiltonian (\ref{eq:H_RWA}) is removed by the latter step. Assuming a time scale separation between the (fast) adaptation of the cavity state to the phase of its driving term and the slow dynamics of the phase $\psi$, two-time perturbation theory can provide a Fokker-Planck equation for the dynamics on slow time scales $r_0 t\; (r_0 \ll 1)$ of the phase distribution probability \cite{Pavbook,BeOrsBook} 
\begin{equation}
    \label{eq:fokker-planck}
    \frac{\partial}{\partial t} P(t,\psi) = -\frac{\partial}{\partial \psi} \big[j(\psi) \, P(t,\psi)\big]+\frac{\partial^2}{\partial \psi^2} \big[D(\psi)\, P(t,\psi)\big].
\end{equation}
The drift coefficient, in the first order of $r_0$, is given by  $j(\psi) = 2\pi r_0\braket{\hat{I}_\mathrm{CP}/(2e)} - \delta_\mathrm{ac}$ 
as derived in \ref{sec: Ap_TTPT}. In second order one obtains a correction for the drift and a diffusion term which are calculated via the pseudo-inverse of the system's Liouvillian.  
Without locking signal we find free diffusion of the phase with diffusion constant, $D = (2\pi r_0)^2 S_\mathrm{CP}/2$, determined by the zero-frequency CP shot noise power $S_\mathrm{CP}$. Injecting the ac-signal creates a potential ${V(\psi) = -\int j(\psi)\; \mathrm{d}\psi}$ for the phase, which develops local minima for sufficient locking strength, restricting diffusion and stabilizing the phase, see figure~\ref{fig:PhaseSlips}(a). For small $\alpha_0$ the potential simplifies to a Shapiro-like tilted washboard $V(\psi) =  \nu_c \cos \psi -\nu_0 \psi$ with $\nu_c = \pi r_0 \varepsilon E_J^*/\hbar$. The diffusion $D$ remains roughly constant in this regime.

The dynamics of a phase particle in this potential is obtained via the Langevin equation corresponding to (\ref{eq:fokker-planck}), see figure~\ref{fig:PhaseSlips}(b). Deep in the synchronized state (green), plateaus of nearly constant phase appear, interrupted by rare slips of the phase by $2\pi$. 
Here, the dynamics may be approximated as an escape process which yields for $\alpha_0\ll 1$ a Kramer's rate $\nu_K = \sqrt{\nu_c^2 - \nu_0^2}\exp\{-\Delta V/[(2\pi r_0)^2 S_\mathrm{CP}/2]\}$ with $\Delta V = 2\sqrt{\nu_c^2 - \nu_0^2} + 2\nu_0 \arcsin(\nu_0/\nu_c) - \pi\nu_0$ \cite{risken1996fokker}. Reducing the strength of the locking drive (orange) accordingly increases the number of slips until the synchronized plateaus are completely absent (blue). How well a system is synchronized, can be quantified by the rate of such slips. From the steady-state distributions for the phase $\psi$, dashed in figure~\ref{fig:PhaseSlips}(a), we obtain the rate of phase slips, i.e., the mean flow $\braket{\dot{\psi}} = \int P(\psi)j(\psi) \mathrm{d}\psi$ in figure~\ref{fig:PhaseSlips}(c). For large $R_0$, the dependence of the phase-slip rate on the locking strength in figure~\ref{fig:PhaseSlips}(c) shows a shot noise induced broadening of the locking transition while in the limit $R_Q/R_0 \to \infty$ shot noise becomes negligible. Then, the Fokker-Planck equation simplifies to the well-known Adler-equation describing the locking dynamics of a classical phase variable, $\dot{\psi}  = \nu_0 + \nu_c \sin(\psi)$\cite{Danner2021}.

\section{Synchronization of two quantum microwave cavities}
\begin{figure}[b]
\includegraphics[width=\textwidth]{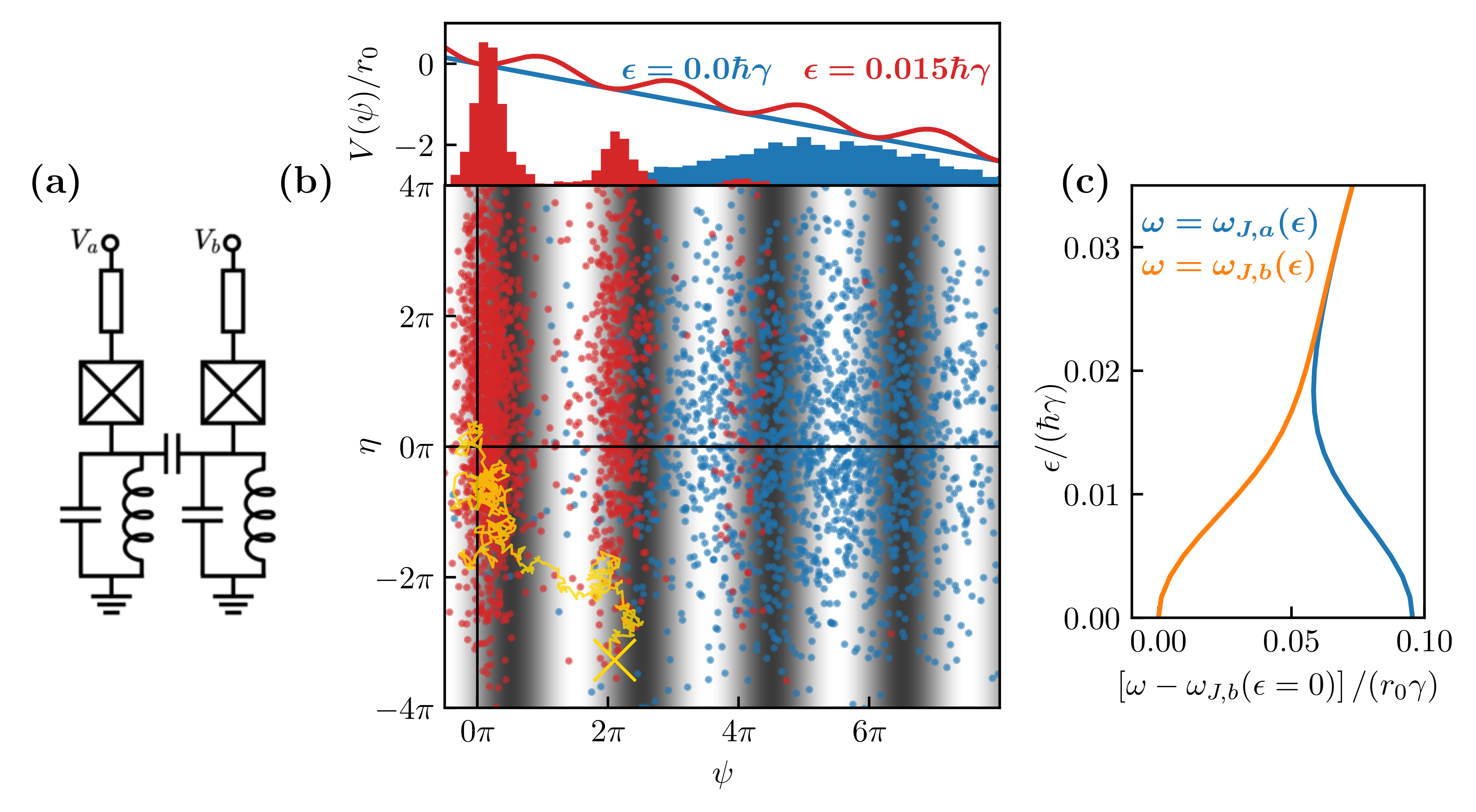}
\caption{Statistics $P(\eta. \psi)$ of the phase-sum and -difference of two Josephson-photonics systems (a). (b) The background color indicates the force on the phase-particle in positive (white) and negative (black) $\psi$-direction respectively which yields the red potential shown in the panel above.
The simulation shows the endpoints of individual trajectories, initially starting at $\eta=\psi=0$, after a time $T=170/(r_0\gamma)$ (one full trajectory is highlighted in yellow with cross at the endpoint). When the oscillators are uncoupled (blue), the phase diffuses freely in both directions. In the synchronized state (red), the phase diffuses freely along $\eta$, while the potential restricts diffusion in $\psi$, resulting in the emergence of two sidebands that indicate phase slips of $2\pi$. (c) Pulling curve of the effective driving frequencies $\omega_J$ coinciding with the respective main emission peaks.
[Parameters: $E_J^*/(\hbar\gamma)=8$, $\alpha_0=0.1$, $2e V_a/\hbar=\omega_0$, $2e(V_a-V_b)/\hbar=0.1 r_0 \gamma$, $r_0 = 1/300$, 2000 trajectories.]}
\label{fig:Synchronization}
\end{figure}

Instead of the synchronization of a single nonlinearly driven microwave cavity to an external locking signal investigated above, one can also consider the mutual synchronization of several devices \cite{Pikovsky2001,Kuramoto1975}. 
Dispensing with the need of a phase-stable ac-signal, the prospect of mutual synchronization of a potentially large number of devices promises a source of correlated high-intensity emission.
The Hamiltonian,
\begin{equation}
H = H_a + H_b + \epsilon (a^{\dagger} b + a b^\dag ), 
\end{equation}
where
\begin{equation}
H_\xi = \hbar \omega_0 \xi^\dag \xi - E_J \cos\!\!\left[\alpha_0 (\xi^\dag + \xi) + \omega_{\mathrm{dc}, \xi} t - 2\pi r_0 m_\xi \right]
\end{equation}
describes a scenario [see Figure~\ref{fig:Synchronization}(a)], where each device $\xi=a,b$ is biased by a dc-voltage $V_\xi = \hbar \omega_{\mathrm{dc}, \xi}/(2e)$ 
applied across a resistance $R_0$, and the coupling term stems from a small coupling capacitor and a rotating wave approximation.
We assume identical devices except for differing dc-voltage drives. 

When two oscillators are in a synchronized state, their relative phase is constant in time. It is therefore useful to introduce sum and difference variables
\begin{equation}
    \eta = -2\pi r_0\Big[(m_a + m_b) - \braket{m_a + m_b}_{\epsilon=0}\Big]
 \end{equation}
\begin{equation}
    \psi = (\omega_{\mathrm{dc}, a} - \omega_{\mathrm{dc}, b}) t - 2\pi r_0(m_a - m_b)\, ,
\end{equation}
so that $\braket{\eta}_{\epsilon=0} = 0$ and $\braket{\psi}_{\epsilon=0}$ drifts with the effective detuning in the uncoupled case.

We again derive a 2d-Fokker-Planck equation for $P(\eta, \psi)$ by using time-scale separation, see \ref{sec:notesynch}. When the systems are uncoupled, the phase particle described by its 2d-position $\vec{x}=(\eta, \psi)$ diffuses freely, as illustrated by the statistics in figure~\ref{fig:Synchronization}.
If the initially detuned oscillators are coupled ($\epsilon>0$) one device will provide a reference frequency $\omega_{J, a}$ for the second device, which pulls the emission frequency of the second cavity $\omega_{J, b}$ towards that reference frequency, and vice versa. In the initially flat potential $V(\eta, \psi)$, minima form along the $\psi$ direction and in the synchronized state both cavities will emit photons about the same frequency, such that $\braket{\dot{\psi}} \approx 0$. Hence, the statistics of the phase difference $\psi$ becomes sharp. However, shot noise permits the phase difference to slip by $2\pi$. Notably, the resulting strong correlation between the cavities is not associated with concomitant entanglement.


\section{Conclusion and outlook}
This paper introduces dc-driven Josephson-photonics devices as a new platform to study quantum synchronization in the presence of shot noise and the thereby induced phase-slip dynamics.
Mechanism and modeling of quantum synchronization dynamics in such devices differ from heretofore investigated systems and do not map to the paradigmatic case of a van der Pol oscillator.
Nonetheless, the system can be reduced to its phase dynamics governed by a Fokker-Planck equation with a generalized Adler potential.
Implementing synchronization in Josephson-photonics devices is of technological interest due their versatility as a source of quantum microwave light, but also promises fundamental insights in a  deep quantum regime, where the nature of phase slips and the potential coexistence and interplay of charge and phase tunneling \cite{Hriscu2013} can be studied.
\section*{Acknowledgments}
We thank Andrew Armour, Julien Basset, Jérôme Estève, Benjamin Huard, and Ambroise Peugeot for fruitful discussions and acknowledge the support of the DFG through AN336/17-1 and AN336/18-1 and the BMBF through QSens (project QComp).

\appendix

\section{Hamiltonian in rotating wave approximation}
\label{sec::Ap_RWA_Ham}

The Hamiltonian of the dc-biased Josephson-photonics device with in-series resistance $R_0$ and an applied voltage ${V =V_\mathrm{dc} + V_\mathrm{ac}\cos(\phi_\mathrm{ac})}$ (where $\phi_\mathrm{ac} = \Omega t + \varphi_\mathrm{ac}$ is the phase of the ac-signal) can be derived from Kirchhoff's laws as
\begin{equation}
    H = \hbar \omega_0 a^\dag a - E_J\cos\left[\omega_\mathrm{dc}t - \varphi_{R_0} + \alpha_0 (a^\dag + a) + \varepsilon \sin(\phi_\mathrm{ac})  \right],
\end{equation}
where $\phi_{a} = \omega_\mathrm{dc} t - \varphi_{R_0}$ (with $\omega_\mathrm{dc}=2eV_\mathrm{dc}/\hbar$) is the driving phase of the cavity, and $\varepsilon=2e V_\mathrm{ac}/(\hbar\Omega)$ is the amplitude of the ac-signal assumed to be small. We move to a reference frame rotating at $\omega_\mathrm{rf} \approx \omega_\mathrm{dc}$ with the unitary operator $U=e^{i\phi_\mathrm{rf} a^\dag a}$, where $\phi_\mathrm{rf} = \omega_\mathrm{rf} t + \phi_0$. The Hamiltonian is then transformed as $H_\mathrm{rf} = UHU^\dag + i\hbar\dot{U}U$. Linearizing the expression in the strength of the ac-signal $\varepsilon$, we find $H_\mathrm{rf} = H_0 + h + h^\dag$ with $H_0 = \hbar (\omega_0-\omega_\mathrm{rf})a^\dag a$ and 
\begin{equation}
        h = - \frac{E_J e^{- \alpha_0^2 / 2}}{2}\Big\{\left[1+i\varepsilon\sin (\phi_\mathrm{ac})\right]e^{i \phi_a} \cdot\,  e^{i\alpha_0 a^\dag \cdot\exp(i\phi_\mathrm{rf}) } \cdot \, e^{- i\alpha_0 a \cdot \exp(-i\phi_\mathrm{rf})}\Big\}\,,\;
\end{equation}
where the Josephson energy is renormalized, $E_J^* = E_J e^{-\alpha_0^2/2}$. Finally, we perform a rotating wave approximation, keeping slowly oscillating terms only, resulting in
\begin{equation}
 \label{eq:A1_hRWA}
 \fl
 \eqalign{
 h \approx -\frac{E_J^* \alpha_0}{2}ie^{-i[\phi_\mathrm{rf} - \phi_a]} :a \frac{J_1(2\sqrt{\alpha_0^2 a^\dag a})}{ \sqrt{\alpha_0^2 a^\dag a}} : \cr
+\varepsilon \frac{E_J^* \alpha_0^2}{8}e^{-i[2 \phi_\mathrm{rf} -\phi_\mathrm{ac} - \phi_a]} : a^2\frac{2 \cdot J_2(2\sqrt{\alpha_0^2 a^\dag a})}{\sqrt{\alpha_0^2 a^\dag a}^2} 
+\varepsilon\frac{E_J^*}{4}e^{-i[\phi_\mathrm{ac} - \phi_a]} J_0(2\alpha\sqrt{a^\dag a}):\,,
}
\end{equation}
where each infinite series of powers of the annihilation and creation operators was written compactly as normally-ordered Bessel-functions. After defining the Adler phase $\psi$ as the phase difference between ac-driving phase and the phase of the cavity $\psi = \phi_\mathrm{ac} - \phi_a $ and choosing the rotating frame such that $\phi_\mathrm{rf} = \phi_\mathrm{ac}$, we find the Hamiltonian (\ref{eq:H_RWA}) from the main text. 



\section{Coherent Cooper pair counting}
\label{sec::Ap_coherent_counting}
The tunneling of Cooper pairs, the process we want to count, appears as part of the coherent time-evolution of the system, i.e., in the Hamiltonian dynamics contribution to the quantum master equation for the density matrix. This contrasts with counting emitted photons or any other incoherent process where there is a single, clear procedure to count: by introducing a counting field in a prefactor $e^{i\chi}$ of the photon jump operator of the Lindblad Master equation to get a generating function from the $\chi$-dependence of the density matrix \cite{Bagrets2003,Novotny2013}, or fully equivalently by working in the corresponding Fourier domain by `number-resolving' the density matrix into components corresponding to a certain number of emitted photons, whose dynamic equations are coupled by photon jumps \cite{Xu2013}. 

Counting is more subtle for coherent processes. In fact, there are various schemes of introducing a counting field to the Hamiltonian part of the dynamics, which have been linked to different (virtual) measurement devices and protocols \cite{Bachmann2010}. Here, we start by attaching $e^{\pm i\chi/2}$ prefactors to all terms in the Hamiltonian corresponding to the forward/backward transfer of a Cooper pair. That Hamiltonian acts from the left side on the density matrix in the coherent part of the time-evolution corresponding to the forward branch of a Keldysh contour \cite{Novotny2013}, while acting from the right and residing on the backward branch, $\chi \rightarrow -\bar{\chi}$ is used. 
For Josephson-photonics setups where each Cooper pair transfer goes along with the creation of a cavity photon (i.e., purely dc-driven biasing without an ac-component) this method (with $\bar{\chi}\equiv\chi$) has been shown \cite{Kubala2020} to yield a coherent counting statistics identical to the incoherent photon emission statistics in the long-time limit.

To capture the effect of transferred charges on the voltage drop at the in-series resistance, namely by the driving phase's dependence on the number of charges, $\varphi_{R_0}=2\pi r_0$, one has to go from $\chi$-space to number space. The number-resolved quantum master equation (\ref{eq:qme}) introduced in the main text corresponds to Fourier-transforming with respect to the sum variable, $\chi+\bar{\chi}$ while summing over the different numbers of transferred charges on both Keldysh branches, so that both, density matrices diagonal in the Keldysh counts (dubbed `classical' in \cite{clerk2016}) and off-diagonal ones contribute to the $\rho_m,\;  2m\in \mathbb{Z}$. The average of transfers on upper and lower branch determines the fed-back charge affecting the driving phase. The off-diagonal contributions have been associated via a picture of interfering paths with negativities making $P(m) = \mathrm{tr}\rho_m$ a quasiprobability at short times which, however, do not matter due to the time-scale separation. It can easily be confirmed, that without the feedback mechanism, i.e., taking $r_0 \rightarrow 0$, the unresolved quantum master equation is found by re-summing and that the Fokker-Planck equation derived from the generalized number-resolved quantum master equation (\ref{eq:qme}) yields the expected current-noise.

\begin{figure}[b]
\centering
\includegraphics[width=\textwidth]{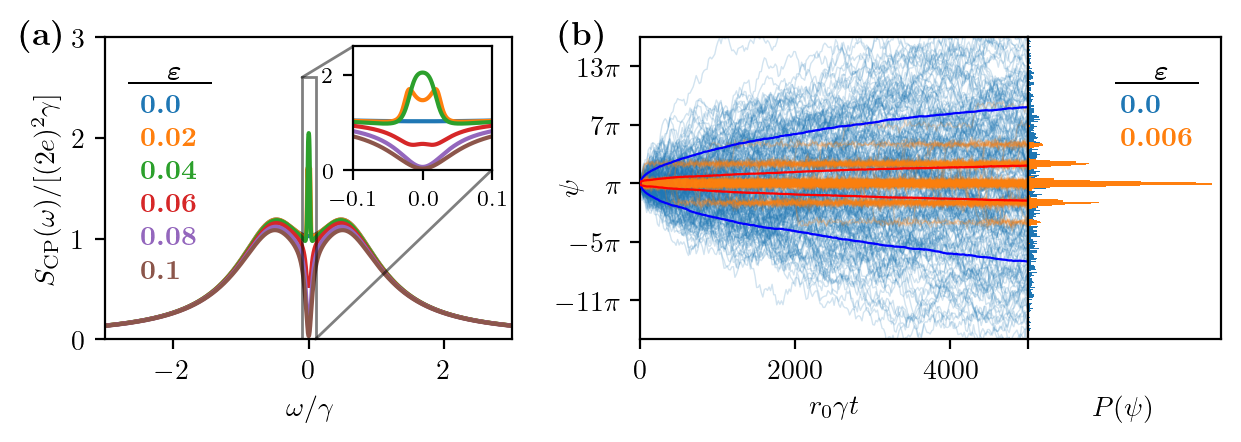}
\caption{(a) The noise of the Cooper-pair current in a single Josephson-photonics device is analyzed for various locking amplitudes $\varepsilon$. At low frequencies, an injection signal initially adds noise for strengths insufficient to achieve locking, but as the system locks for larger $\varepsilon$, the zero-frequency noise becomes almost entirely suppressed. [Parameters see figure~\ref{fig:SpectrumWaterfall}.] (b) The long-time full counting statistic of CP transport is calculated from the distribution of the phase $\psi \in \mathbb{R}$ gained from simulating 2000 trajectories. The standard deviation $\Delta \psi$ shows the $\propto \sqrt{t}$ behavior typical for diffusive processes, where the free diffusion (blue) is replaced by a much slower phase-slip process in the (weakly) locked case. 
 [Parameters: $E_J^*/(\hbar\gamma)=10$, $\alpha_0=0.1$, $2e V_\mathrm{dc}/\hbar=\omega_0$, $\nu_0 = 0$, $r_0 = 1/300$.]}
\label{fig:B}
\end{figure}

The \emph{frequency-dependent current noise}
\begin{equation}
    S_\mathrm{CP}(\omega) = 2\mathrm{Re} \int_0^\infty \mathrm{d}\tau \braket{\delta I(t+\tau)\delta I(t)}e^{i\omega \tau}
\end{equation}
in the steady-state ($t\to\infty$) with $\delta I = I_\mathrm{CP} - \braket{I_\mathrm{CP}}$ is calculated from the full dynamical equations (\ref{eq:H_RWA}) and (\ref{eq:qme}) shown in figure \ref{fig:B}(a).
Without locking the noise is the same as for the idealized case of a fixed driving phase. Due to the nonlinearity of the Josephson-photonics system the transport is slightly sub-Poissonian \cite{Armour2017} and the noise shows (fast) features on the scale of $\gamma$ in this regime. As the ac-locking signal is switched on, the slow phase dynamics changes from free diffusion to the drift and diffusion in a periodic potential and the current-correlations show signatures of this motion as sidepeaks at a frequency $|\omega| \lesssim |\nu_0|$ (yellow), which merge and develop into a dip as the phase gets trapped in the potential's minima. The feedback of the transferred charges on the driving phase correlates current fluctuations over very long times and correspondingly suppresses the zero-frequency noise down to a value connected to the total rate of phase slips. On time scales faster than the phase dynamics, the current fluctuations remain nearly unaffected. 

The zero-frequency noise is of course linked to the second cumulant of the CP counting statistics. The long-time limit of $P(m)$ shown in figure \ref{fig:SpectrumWaterfall} (c) of the main text can be more easily calculated from the reduced phase dynamics, so that $P(\psi)$ (with $\psi \in \mathbb{R}$) gained from a histogram of 2000 Langevin trajectories is shown here in \ref{fig:B}(b). Without locking the full counting statistics spreads out via free phase diffusion corresponding to (near-)Poissonian Cooper pair transport events as $\langle\langle \psi \rangle\rangle \propto \langle\langle m \rangle\rangle \propto D t$, while in the locked regime $\psi$ is distributed in peaks around the minima of the potential and spreads very slowly between these peaks by phase-slip processes.

\section{Derivation of the reduced equation for the phase}
\label{sec: Ap_TTPT}
Starting from the full dynamical equations we first write (\ref{eq:qme}) in terms of the phase $\varphi_{R_0}=2\pi r_0 m$ instead of the number $m$ of transferred CP and treat $\varphi_{R_0}$ as a continuous quantity ${\rho_m(t) \to \rho(t, \varphi_{R_0} = 2\pi r_0 m)}$. 
We want to express $\varphi_{R_0}$ in terms of a phase $\psi$ in a moving reference frame, which is defined by the transformation, cf. (\ref{eq:psi}),
\begin{equation}
    \varphi_{R_0}(t, \psi) = \psi + \omega_\mathrm{dc}t - (\Omega t + \varphi_\mathrm{ac}).
\end{equation}
Accordingly, the density matrix becomes a new function of $t$ and $\psi$
\begin{equation}
    \rho'(t, \psi) = \rho\big(t, \varphi_{R_0}(t, \psi)\big)
\end{equation}
with the time derivative
\begin{equation}
    \label{eq:raw_derivative}
    \fl
    \frac{\partial \rho'}{\partial t} (t, \psi) = \frac{\mathrm{d}\rho}{\mathrm{d}t}\big(t, \varphi_{R_0}(t, \psi)\big) = \frac{\partial\rho}{\partial t} \big(t, \varphi_{R_0}(t, \psi)\big) + \frac{\partial\rho}{\partial \varphi_{R_0}}\big(t, \varphi_{R_0}(t, \psi)\big)\frac{\partial \varphi_{R_0}}{\partial t}(t, \psi).
\end{equation}
Introducing the evolution generator $\Lambda(\psi)$ this yields
\begin{equation}
    \label{eq:cont_propagator}
    \fl
    \eqalign{\frac{\partial\rho'}{\partial t} (t, \psi) = \Lambda \rho'(\psi) = \delta_\mathrm{ac}\frac{\partial\rho'}{\partial \psi} (t, \psi) + \mathcal{L}_0 \rho'(\psi)  -\frac{i}{\hbar}\Big[h^\dag(\psi-\pi r_0) \rho'(\psi-\pi r_0) \cr 
    + h(\psi+\pi r_0) \rho'(\psi+\pi r_0) - \rho'(\psi+\pi r_0)h^\dag(\psi+\pi r_0) - 
    \rho'(\psi-\pi r_0)h(\psi-\pi r_0)\Big]}
\end{equation}
Note that  the Hamiltonian $h(\psi)=h(t, \varphi_{R_0}(t, \psi))$ does not have an explicit time dependence, see (\ref{eq:H_RWA}).\\

We perform two time-scale perturbation theory (or homogenization) \cite{pavliotis2008multiscale, DoMu2019} to derive an effective Fokker-Plank equation for the phase $\psi$. For that purpose it is assumed that the oscillator dynamics takes place on a much faster time scale than the phase dynamics. 
We explicitly introduce a fast time scale $t$ and a slow time scale $\tau=r_0 t$ and the dependence of the state on these times $\rho'(t, \psi)=\rho(t,\tau, \psi)$.  Then, the derivative of the density matrix becomes 
\begin{eqnarray}
    \partial_t \rho'(t, \psi) &=\frac{\mathrm{d}}{\mathrm{d}t}\rho(t, \tau, \psi) = \partial_t\rho(t,\tau, \psi)+(\partial_t \tau)\partial_\tau\rho(t,\tau, \psi) \nonumber \\
    &= \partial_t\rho(t,\tau, \psi)+ r_0\partial_\tau\rho(t,\tau, \psi).    
\end{eqnarray}
Furthermore, we assume that the fast oscillator dynamics have already relaxed and we study the evolution of the state $\rho(\tau, \psi)= \lim_{t\rightarrow\infty}\rho(t,\tau, \psi)$, so that (\ref{eq:cont_propagator}) becomes
\begin{equation}
    \label{eq:taupart}
    r_0\partial_\tau\rho(\tau, \psi) = \Lambda \rho(\tau, \psi)
\end{equation}
In a next step, we expand $\rho(\tau, \psi) = \sum_{l=0}^\infty r_0^l \rho^{(l)}$ and $\Lambda \rho = \mathcal{L}\rho + r_0\partial_\psi( \Lambda^{(1)}\rho) + \frac{1}{2}r_0^2 \partial_\psi^2(\Lambda^{(2)}\rho) + \mathcal{O}(r_0^3)$ in orders of $r_0$. 
Taking $ \rho(\psi \pm \pi r_0) = \rho \pm \pi r_0 \partial_\psi \rho + \mathcal{O}(r_0^2)$ and $h(\psi \pm \pi r_0) = h \pm \pi r_0 \partial_\psi \rho + \mathcal{O}(r_0^2)$ results in the zeroth-order Liouvillian $\mathcal{L}\rho = -\frac{i}{\hbar} [h + h^\dag, \rho] + \mathcal{L}_0\rho$ and for the next order $\Lambda^{(1)} = \delta_\mathrm{ac}/r_0 - 2\pi \mathcal{I}_\mathrm{CP}/(2e)$ with the symmetrized current operator $\mathcal{I}_\mathrm{CP}\rho = \frac{1}{2}(I_\mathrm{CP}\rho + \rho I_\mathrm{CP})$.

We proceed by solving (\ref{eq:taupart}) order by order. In \emph{zeroth order} or $r_0$ we obtain 
\begin{equation}
    0 = \mathcal{L}(\psi)\rho^{(0)}(\tau, \psi)
\end{equation}
which determines $\rho^{(0)}(\tau, \psi)=P^{(0)}(\tau, \psi)\rho_\mathrm{eq}(\psi)$ up to a factor which we interpret as a probability density $P^{(0)}(\tau, \psi)$ for a certain $\psi$ at time $\tau$. To find it, we consider the  \emph{first order} of $\rho(\tau, \psi)$ in $r_0$
\begin{equation}
    \partial_\tau \rho^{(0)}(\tau, \psi) = \mathcal{L}(\psi) \rho^{(1)}(\tau, \psi) + \partial_\psi\left\{\left[\delta_\mathrm{ac}/r_0 -2\pi \mathcal{I}_\mathrm{CP}(\psi)/(2e)\right] \rho^{(0)}(\tau, \psi)\right\}.
\end{equation}
Taking the trace yields the desired equation of motion
\begin{equation}
    \partial_\tau P^{(0)}(\tau, \psi) = \partial_\psi \left[ \braket{\delta_\mathrm{ac}/r_0 -2\pi \mathcal{I}_\mathrm{CP}(\psi)/(2e)}_{\rho_\mathrm{eq}(\psi)}P^{(0)}(\tau, \psi)\right]
\end{equation}
which describes a drift of $\psi$. The \emph{second order} (see supplemental material) in $r_0$ provides a correction for the drift and a diffusion term.\\

Then the equation of motion can be written as the Fokker-Planck equation in (\ref{eq:fokker-planck})
with drift
\begin{equation}
    \fl
    j(\psi) = 2\pi r_0\braket{\mathcal{I}_\mathrm{CP}/(2e)}_{\rho_\mathrm{eq}(\psi)} - \delta_\mathrm{ac}
    +r_0^2\braket{\left(\partial_\psi2\pi \mathcal{I}_\mathrm{CP}/(2e)\mathcal{R}\right)\left(\delta_\mathrm{ac}/r_0 - 2\pi  \mathcal{I}_\mathrm{CP}/(2e)\right)}_{\rho_\mathrm{eq}(\psi)}
\end{equation}
and diffusion
\begin{equation}
    D(\psi) = - (2\pi r_0)^2\braket{\mathcal{I}_\mathrm{CP}/(2e) \mathcal{R} \mathcal{I}_\mathrm{CP}/(2e)}_{\rho_\mathrm{eq}(\psi)}
\end{equation}
where $\mathcal{R}$ is the pseudo-inverse of $\mathcal{L}$. The $r_0^2$ contribution in the drift term may be neglected for small $r_0$.

\section{Note on mutual synchronization}
\label{sec:notesynch}
Two Josephson-photonics devices can be coupled, e.\,g.\,, inductively or capacitively, where we here choose to analyze a capacitive coupling. Since charge and phase are conjugate variables associated with momentum and position of the cavities, a capacitive coupling will lead to a interaction term of the form $\hat{Q}_a \hat{Q}_b \propto (a^\dag - a) (b^\dag - b)$ in the Hamiltonian. A sufficiently small capacitor will lead in RWA to the Hamiltonian given in the main text. We further note that a coupling capacitor modifies the eigenfrequencies of the cavities. 
 
As in the case of locking to an ac-signal, a Fokker-Planck equation can also be derived for the synchronization of two devices. Similar to the case of injection locking, one introduces new phase variables $\varphi_a'$ and $\varphi_b'$ which are given by $\varphi_a' = -(\eta + \psi)/2$ and $\varphi_b = -(\eta-\psi)/2$. Calculations analogous to  \ref{sec: Ap_TTPT} yield a Fokker-Planck equation for the probability density $P(t, \varphi_a, \varphi_b)$
 \begin{equation}
    \fl
     \frac{\partial}{\partial t} P = \left\{-\frac{\partial}{\partial \varphi_a'} j_a -\frac{\partial}{\partial \varphi_b'} j_b + \frac{\partial^2}{\partial \varphi_a'^2}D_{aa} + 2\frac{\partial^2}{\partial \varphi_a' \partial \varphi_b'} D_{ab}  + \frac{\partial^2}{\partial \varphi_b'^2} D_{bb}\right\}P
 \end{equation}
with drift and diffusion coefficients given in the supplemental material.

\section*{References}
\bibliographystyle{unsrt}
\bibliography{referencesQS_nomonth}
\end{document}


\title[Supplemental Material]{Supplemental Material to\\Quantum Synchronization in Presence of Shot Noise}

\author{
Florian H\"ohe$^1$,
Lukas Danner$^{1,2}$, 
Ciprian Padurariu$^1$,
Brecht I. C Donvil$^1$, 
Joachim Ankerhold$^1$, and
Bj\"orn Kubala$^{1,2}$
}

\address{$^1$Institute for Complex Quantum Systems and IQST, University of Ulm, 89069 Ulm, Germany}
\address{$^2$Institute for Quantum Technologies, German Aerospace Center (DLR), 89081 Ulm, Germany}
\ead{florian.hoehe@uni-ulm.de}
\vspace{10pt}

%
%

In the supplemental material we complete the derivation of the equation for the reduced phase of Appendix C. Furthermore, we give details on Kramer's escape rate which approximates the phase dynamics and we present the reduced equation for the phases of two synchronized Josephson-photonics devices.
\section{Second order correction to the Fokker-Planck equation}
\begin{figure}[b]
\includegraphics[width=\textwidth]{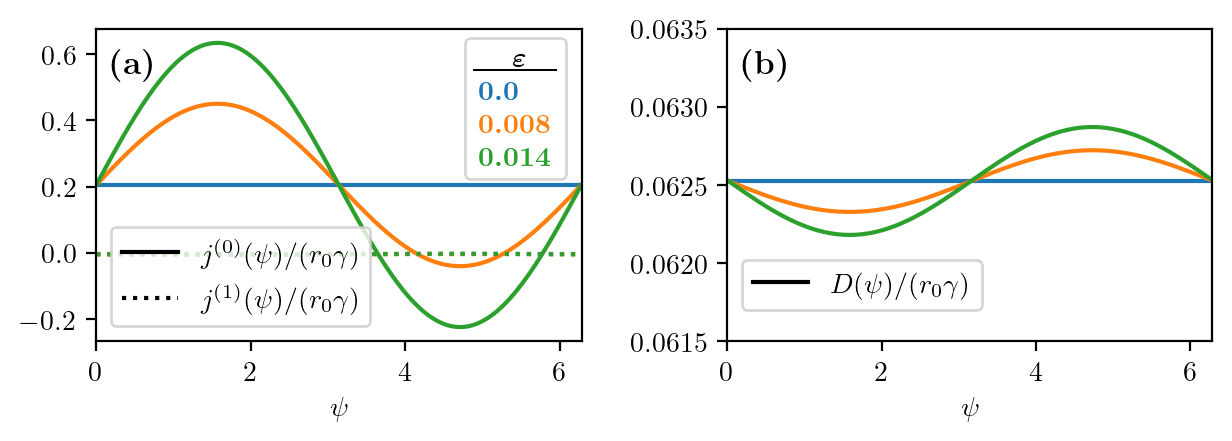}
\caption{Drift and diffusion coefficients for injection locking. (a) The drift coefficient shows a strong phase dependence for $\varepsilon>0$ while the diffusion coefficients (b) only show a weak dependence. [Parameters see figure 3 in the main text.]}
\label{fig:drift_diff_coeff}
\end{figure}

First we note, that by taking the trace of the first order equation
\begin{equation}
    \partial_\tau \sket{\rho^{(0)}(\tau, \psi)} = \mathcal{L}(\psi) \sket{\rho^{(1)}(\tau, \psi)} + \partial_\psi\left[\Lambda^{(1)}(\psi) \sket{\rho^{(0)}(\tau, \psi)}\right].
\end{equation}
we did not extract its full information. To do so, let $\mathcal{R}$ be the pseudoinverse (Moore–Penrose inverse) of $\mathcal{L}$ i.e. $\mathcal{R} = (\mathcal{L} + \mathcal{P})^{-1} - \mathcal{P}$ and with $\mathcal{P} = \sket{\rho_\mathrm{eq}}\sbra{1}$
\begin{equation}
    \mathcal{L}\mathcal{R} = \mathcal{R}\mathcal{L} = \mathcal{Q}
\end{equation}
applies, where $\mathcal{Q} = 1 - \mathcal{P}$. Then multiplying with $\mathcal{R}$ from left yields
\begin{equation}
    0 = \mathcal{R}\mathcal{L}\sket{\rho^{(1)}} + \mathcal{R}\partial_\psi\left(\Lambda^{(1)}\sket{\rho^{(0)}}\right)
\end{equation}
and equivalently
\begin{equation}
    \label{eq:uselater}
    \mathcal{Q}\sket{\rho^{(1)}} = - \mathcal{R}\partial_\psi\left(\Lambda^{(1)}\sket{\rho^{(0)}}\right).
\end{equation}
The second order equation reads
\begin{equation}
        \partial_\tau \sket{\rho^{(1)}} = \mathcal{L} \sket{\rho^{(2)}} + \partial_\psi\left(\Lambda^{(1)}\sket{\rho^{(1)}} \right) +\frac{1}{2}\partial_\psi^2 \left(\Lambda^{(2)}\sket{\rho^{(0)}}\right).
\end{equation}
From the the expansion of $\Lambda$ in $r_0$ we find 
\begin{equation}
    \Lambda^{(2)}\rho = -i\pi^2 r_0^2 [h^\dag + h, \rho].
\end{equation}
Taking the trace, using $\mathcal{Q}+\mathcal{P} = 1$, and inserting (\ref{eq:uselater}) gives
\begin{eqnarray}
    \fl
    \eqalign{\partial_\tau P^{(1)} =& \phantom{+} \partial_\psi \sbra{1}\partial_\psi\left(\Lambda^{(1)}\mathcal{R}\right)\Lambda^{(1)}\sket{\rho^{(0)}} + \partial_\psi \left(\sbra{1}\Lambda^{(1)}\sket{\rho_\mathrm{eq}}P^{(1)}\right)\cr
    &+\frac{1}{2}\partial_\psi^2\sbra{1}\Lambda^{(2)}\sket{\rho^{(0)}}-\partial_\psi^2 \sbra{1}\Lambda^{(1)}\mathcal{R}\partial_\psi\Lambda^{(1)}\sket{\rho^{(0)}}}
\end{eqnarray}
where we made use of $A\partial_x(B)=\partial_x(AB)-\partial_x(A)B$.  As $\Lambda^{(2)}$ is a commutator with $\rho$ its expectation value vanishes. 
Finally we approximate ${P \approx P^{(0)} + P^{(1)}}$. We expect the $P^{(1)}$ contribution to the other terms will come in the next order of $r_0$ and already add it. With that we obtain the Fokker-Planck equation
\begin{equation}
    \partial_\tau P(\tau, \psi) = -\partial_\psi \left[(j^{(0)}(\psi) + j^{(1)}(\psi)) P(\tau, \psi)\right] + \partial_\psi^2\left[D(\psi) P(\tau, \psi)\right]
\end{equation}
with
\begin{equation}
    \label{eq:j0}
    j^{(0)}(\psi) = -\sbra{1}\Lambda^{(1)}\sket{\rho_\mathrm{eq}}
\end{equation}
\begin{equation}
    \label{eq:j1}
    j^{(1)}(\psi) = -\sbra{1}\partial_\psi\left(\Lambda^{(1)}\mathcal{R}\right) \Lambda^{(1)}\sket{\rho_\mathrm{eq}}
\end{equation}
\begin{equation}
    \label{eq:D}
    D(\psi) = - \sbra{1}\Lambda^{(1)} \mathcal{R} \Lambda^{(1)}\sket{\rho_\mathrm{eq}}.
\end{equation}
 and we define $j(\psi) = j^{(0)}(\psi) + j^{(1)}(\psi)$. We show the different contributions to the drift, (\ref{eq:j0}) and (\ref{eq:j1}), and the diffusion term (\ref{eq:D}) in figure \ref{fig:drift_diff_coeff}.\\

As described in the main text, the potential is determined by the drift term, $V(\psi) = - \int j(\psi)\, \mathrm{d}\psi$, which in turn is gained from the expectation value of the dc-current $I_{\mathrm{CP}}$ in the steady state of $\mathcal{L}(\psi)$. In the limit of small zero-point fluctuations, the Hamiltonian and the current operator can be approximated, only including bilinear operator terms,
\begin{equation}
\fl
H_\mathrm{RWA}
= \left[ \hbar \Delta - \varepsilon E_J^* \frac{\alpha_0^2}{2}  \cos(\psi)\right] a^\dag a +  \frac{E_J^*}{2}\left(i e^{i \psi} \alpha_0 a^\dag 
   + e^{i\psi}\varepsilon\frac{\alpha_0^2}{4}  a^{\dag 2} + h.c.\right) 
\end{equation}
\begin{equation}
  \fl
  \frac{I_\mathrm{CP}}{2e} 
  =\varepsilon\frac{E_J^*}{2\hbar} \sin(\psi)  \left(1 - \alpha_0^2 a^\dag a \right)
  +\frac{E_J^*}{2\hbar}  \left( e^{i \psi}\alpha_0 a^\dag -i e^{i\psi}\varepsilon\frac{\alpha_0^2}{4} a^{\dag 2} + h.c.\right)    
\end{equation}

For this bilinear problem we can find a closed set of coupled equations for expectation values in the steady state
\begin{equation}
  \fl\left[\frac{\hbar \gamma}{2} + i \left(\hbar\Delta - \frac{E_J^* }{2} \varepsilon \alpha_0^2 \cos{\psi}\right)\right] \braket{a} + i\frac{E_J^*}{4} \alpha_0^2 \varepsilon e^{i\psi} \braket{a^\dag} = \frac{E_J^*}{2} e^{i\psi}
\end{equation}
\begin{equation}
  \fl\left[\hbar \gamma + 2i \left(\hbar\Delta-  \frac{E_J^*}{2}\varepsilon \alpha_0^2 \cos{\psi}\right)\right] \braket{a^2} -E_J^*\alpha_0 e^{i\psi}\braket{a} + i \frac{E_J^* }{2} \alpha_0^2 \varepsilon e^{i\psi}\braket{a^\dag a} = -i \frac{E_J^*}{4} \alpha_0^2\varepsilon e^{i\psi}      
\end{equation}
\begin{equation}
  \fl \hbar \gamma\braket{a^\dag a} - \frac{E_J^*}{2} \alpha_0 \left(e^{i\psi}\braket{a^\dag} + e^{-i\psi} \braket{a}\right) +  \frac{E_J^* }{4}\alpha_0^2 \varepsilon \left(ie^{i\psi}\braket{a^{\dag 2}} -ie^{-i\psi}\braket{a^2}\right) =  0\; . 
\end{equation}

 In linear order of $\varepsilon$, we find
 \begin{equation}
     \braket{a} \approx \frac{E_J^*\alpha_0 }{\hbar \gamma} e^{i \psi} + \frac{i}{2} \varepsilon \left(\frac{E_J^* \alpha_0}{\hbar \gamma}\right)^2 \alpha_0 e^{2i\psi}
 \end{equation}
 \begin{equation}
     \braket{a^2} \approx  \left(\frac{E_J^* \alpha_0}{\hbar \gamma}\right)^2 e^{2i\psi} - \frac{i}{4} \varepsilon \alpha_0 \frac{E_J^* \alpha_0}{\hbar \gamma} 
 \end{equation}
 \begin{equation}
     \braket{a^\dag a} \approx \left(\frac{E_J^* \alpha_0}{\hbar \gamma}\right)^2 - \varepsilon \alpha_0 \left(\frac{E_J^* \alpha_0}{\hbar \gamma}\right)^3 
 \end{equation}
 Thus, the steady state current is 
 \begin{equation}
   \frac{\braket{I_\mathrm{CP}}}{2e\gamma}(\psi) \approx \frac{E_J^*}{\hbar \gamma} \frac{\varepsilon}{2} \sin(\psi) + \alpha_0^2 \left(\frac{E_J^*}{\hbar \gamma}\right)^2.
 \end{equation}
For this limiting scenario we thus have explicitly found an analytical expression for the $\psi$-dependence of the current. In addition to the conventional ($\psi$-independent) Josephson photonics contribution, the ac-signal causes a Shapiro-like term, which leads to locking.

\section{Kramer's escape rate}
\begin{figure}[b]
\includegraphics[width=0.5\textwidth]{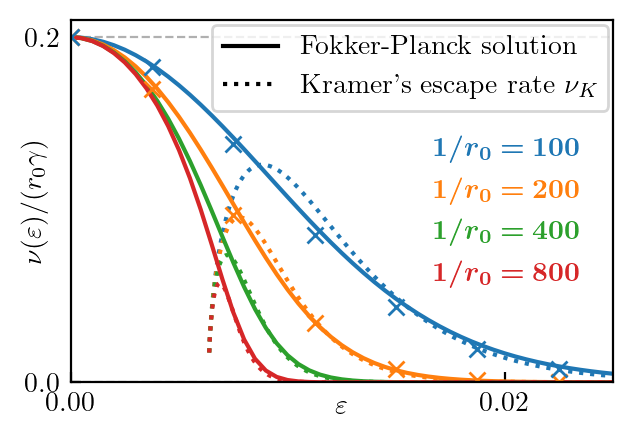}
\caption{Comparison of two-time perturbation theory and Kramer's escape rate with the numerical solution of the master equation. The smaller the resistance $r_0$, the better the Fokker-Planck solution fits the numerical result (crosses). Kramer's escape rate (dotted) is calculated with a diffusion constant extracted from the two-timing result and only matches the Fokker-Planck solution for a clearly pronounced minimum. [Parameters: $E_J^*/(\hbar\gamma)=10$, $\alpha_0=0.1$, $2eV_\mathrm{dc}/\hbar=\omega_0$, $\nu_0 = 0.2 r_0\gamma$.]}
\label{fig:kramer}
\end{figure}

In angular frequencies Kramer's escape rate $r_K$ is given by \cite{risken1996fokker}
\begin{equation}
    \fl
    \nu_K = 2\pi r_K = \sqrt{V''(\psi_\mathrm{min})|V''(\psi_\mathrm{max})|}\exp\left[-(V(\psi_\mathrm{max}) - V(\psi_\mathrm{min}))/D\right]\;,
\end{equation}
where we use that for small $r_0$ the coefficient $D(\psi)$ is only weakly $\psi$-dependent $D(\psi)\approx \mathrm{const.} = D$, cf. figure \ref{fig:drift_diff_coeff}(b). For a sinusoidal potential
\begin{equation}
    V(\psi) = \nu_c \cos \psi -\nu_0 \psi
\end{equation}
we find
\begin{equation}
    \psi_\mathrm{min} = \pi + \arcsin\frac{\nu_0}{\nu_c} \qquad \mathrm{and} \qquad \psi_\mathrm{max}=3\pi-\arcsin\frac{\nu_0}{\nu_c}
\end{equation}
and with that the rate given in the main text. This is a good approximation for tilted potentials with pronounced minima ($\varepsilon$ large enough to be well locked), as shown in figure \ref{fig:kramer} of the main text.

\section{Derivation of the reduced equation for mutual synchronization}
The reduced equation for the phases of two coupled Josephson-photonics devices is derived in a similar way as in the case for injection locking of a single device. 
First, the Hamiltonian (\hamil) in the main text is transformed to a frame rotating with the same frequency $\omega_r$ for $a$ and $b$ via the unitary transformation $U=e^{i\omega_r t (a^\dag a + b^\dag b)}$ such that
\begin{equation}
    \omega_{\mathrm{dc}, a} = \omega_r + \delta_a \qquad \mathrm{and} \qquad \omega_{\mathrm{dc}, b} = \omega_r + \delta_b
\end{equation}
to obtain 
\begin{eqnarray}
    H = &\hbar(\omega_a - \omega_{\mathrm{dc}, a} + \delta_a)a^\dag a + \hbar(\omega_a - \omega_{\mathrm{dc}, a} + \delta_a)a^\dag a \nonumber \\ 
    &+ (h_a^\dag + h_a) + (h_b^\dag + h_b) +\epsilon (a^\dag b + a b^\dag)
\end{eqnarray}
with
\begin{equation}
    h_a^\dag = i\frac{E_J^*}{2}:e^{-i(\delta_a - \varphi_a)}a^\dag \frac{J_1(2\alpha_0 \sqrt{a^\dag a})}{\sqrt{a^\dag a}}:
\end{equation}
and analogous for $b$. Note that the coupling term remains time-independent.\\

We again utilize Cooper-pair counting as in (\cprme) of the main text, but count CPs transferred across the resistors in system $a$ and $b$ separately. Hence, the dynamics of the density matrix $\rho_{m_a, m_b}$ is governed by
\begin{eqnarray}
    \label{eq:qme}
    \fl
    \dot{\rho}_{m_a, m_b} =  \mathcal{L}_0 \rho_{m_a, m_b} \nonumber\\ 
    \fl \phantom{\dot{\rho}_{m_a, m_b} =}
    -\frac{i}{\hbar}\big[h_a^\dag(m_a-1/2, m_b) \rho_{m_a-1/2, m_b}+ h_a(m_a+1/2, m_b) \rho_{m_a+1/2, m_b} \big] + h.c. \nonumber \\
    \fl \phantom{\dot{\rho}_{m_a, m_b} =}
    -\frac{i}{\hbar}\big[h_b^\dag(m_a, m_b-1/2) \rho_{m_a, m_b-1/2}+ h_b(m_a, m_b+1/2) \rho_{m_a, m_b+1/2} \big] + h.c.\;.
\end{eqnarray}
Again, we treat $\varphi_a = 2\pi r_0m_a$ and $\varphi_b = 2\pi r_0m_b$ as continuous and transform to a moving reference frame 
\begin{equation}
    \varphi_a' = \varphi_a - \delta_a t \qquad \mathrm{and} \qquad \varphi_b' = \varphi_b - \delta_b t.
\end{equation}
Hereby, the Hamiltonian becomes time-independent. We choose $\delta_a$ and $\delta_b$ such that they are of the order of $r_0 \gamma$. After expanding in $r_0$ and solving the resulting equation order by order, we end up with a two-dimensional Fokker-Planck equation
 \begin{equation}
    \fl
     \frac{\partial}{\partial t} P = \left\{-\frac{\partial}{\partial \varphi_a'} j_a -\frac{\partial}{\partial \varphi_b'} j_b + \frac{\partial^2}{\partial \varphi_a'^2}D_{aa} + 2\frac{\partial^2}{\partial \varphi_a' \partial \varphi_b'} D_{ab}  + \frac{\partial^2}{\partial \varphi_b'^2} D_{bb}\right\}P
 \end{equation}
with drift and diffusion coefficients
\begin{equation}
    j_a^{(0)}(\varphi_a', \varphi_b') = -\sbra{1}\Lambda_a\sket{\rho_\mathrm{eq}}
\end{equation}
\begin{equation}
    j_b^{(0)}(\varphi_a', \varphi_b') = -\sbra{1}\Lambda_b\sket{\rho_\mathrm{eq}}
\end{equation}
\begin{equation}
    j_a^{(1)}(\varphi_a', \varphi_b') = -\sbra{1}\partial_{\varphi_a'}\left(\Lambda_a\mathcal{R}\right)\Lambda_a\sket{\rho_\mathrm{eq}} -\sbra{1}\partial_{\varphi_b'}\left(\Lambda_a\mathcal{R}\right)\Lambda_b\sket{\rho_\mathrm{eq}}
\end{equation}
\begin{equation}
    j_b^{(1)}(\varphi_a', \varphi_b') = -\sbra{1}\partial_{\varphi_a'}\left(\Lambda_b\mathcal{R}\right)\Lambda_a\sket{\rho_\mathrm{eq}} -\sbra{1}\partial_{\varphi_b'}\left(\Lambda_b\mathcal{R}\right)\Lambda_b\sket{\rho_\mathrm{eq}}
\end{equation}
\begin{equation}
    D_{aa}(\varphi_a', \varphi_b') = -\sbra{1}\Lambda_a\mathcal{R}\Lambda_a\sket{\rho_\mathrm{eq}}
\end{equation}
\begin{equation}
    D_{ab}(\varphi_a', \varphi_b') = -\sbra{1}\Lambda_a\mathcal{R}\Lambda_b\sket{\rho_\mathrm{eq}} - \sbra{1}\Lambda_b\mathcal{R}\Lambda_a\sket{\rho_\mathrm{eq}}
\end{equation}
\begin{equation}
    D_{bb}(\varphi_a', \varphi_b') = -\sbra{1}\Lambda_b\mathcal{R}\Lambda_b\sket{\rho_\mathrm{eq}}\;,
\end{equation}
where
\begin{equation}
    \Lambda_a = -2\pi r_0\mathcal{I}_{\mathrm{CP}, a}/(2e) +  \delta_a 
\end{equation}
\begin{equation}
    \Lambda_b = -2\pi r_0\tilde{R}_0\mathcal{I}_{\mathrm{CP}, b}/(2e) +  \delta_b.
\end{equation}
Here, $\mathcal{I}_{\mathrm{CP}, a}$ and $\mathcal{I}_{\mathrm{CP}, b}$ are the current operators for system $a$ and $b$ respectively. As before $\mathcal{R}$ is the pseudo-inverse of the Liouvillian $\mathcal{L} = -\frac{i}{\hbar}[H, \rho] + \gamma \mathcal{D}[a]\rho + \gamma \mathcal{D}[b]\rho$. A transformation to sum and difference variables $\eta = -(\varphi_a' + \varphi_b')$ and $\psi=-(\varphi_a' - \varphi_b')$ is useful as the drift and diffusion coefficients are independent of $\eta$.

\section*{References}
\bibliographystyle{unsrt}
\bibliography{referencesQS.bib}